\documentclass{article}
\usepackage[preprint]{neurips_2023}
\usepackage[utf8]{inputenc} 
\usepackage[T1]{fontenc}    
\usepackage{hyperref}       
\usepackage{url}            
\usepackage{booktabs}       
\usepackage{amsfonts}       
\usepackage{amsmath}
\usepackage{makecell}
\usepackage{nicefrac}       
\usepackage{microtype}      
\usepackage{verbatim}
\usepackage{courier}
\usepackage{float}
\usepackage{graphicx}
\usepackage{subcaption}
\usepackage[table,xcdraw]{xcolor}
\usepackage{alltt}
\usepackage{tikz}

\makeatletter
\renewcommand{\@notice}{}
\makeatother

\title{A Case Study of Web App Coding with OpenAI Reasoning Models}
\author{%
	Yi Cui \\
	ONEKQ Lab\\
	{yi@onekq.ai} \\
}

\begin{document}	
\maketitle

\begin{abstract}
This paper presents a case study of coding tasks by the latest reasoning models of OpenAI, i.e. o1-preview and o1-mini, in comparison with other frontier models. The o1 models deliver SOTA results for WebApp1K, a single-task benchmark. To this end, we introduce WebApp1K-Duo, a harder benchmark doubling number of tasks and test cases. The new benchmark causes the o1 model performances to decline significantly, falling behind Claude 3.5. Moreover, they consistently fail when confronted with atypical yet correct test cases, a trap non-reasoning models occasionally avoid. We hypothesize that the performance variability is due to instruction comprehension. Specifically, the reasoning mechanism boosts performance when all expectations are captured, meanwhile exacerbates errors when key expectations are missed, potentially impacted by input lengths. As such, we argue that the coding success of reasoning models hinges on the top-notch base model and SFT to ensure meticulous adherence to instructions.
\end{abstract}

\section{Introduction}
The recent release of OpenAI reasoning models (o1-preview and o1-mini)\citep{openai-o1} presents a groundbreaking direction for model development, along with their SOTA performance in several challenging benchmarks, including math\citep{aime}, scientific research\citep{gpqa}, competitive programming\citep{codeforces}.

In this report, we evaluate o1 models in the context of practical software development, i.e. when models are required to implement simple web apps satisfying specific requirement\citep{webapp1k-paper}. Our benchmarks have the following characteristics and challenges.
\begin{itemize}
\item The problem is \textit{less explorational and more results-oriented} than other benchmarks. The specific instructions are laid out in the form of test setup and expectations.
\item \textit{No external knowledge} is required to complete the task, since React is a prominent framwork with sufficient code circulating on Internet for a decade.
\item Some expectations are \textit{less explicit or less typical} than others, which could cause model negligence or misunderstanding.
\end{itemize}

We use a single-task benchmark (WebApp1K) and a duo-task benchmark (WebApp1K-Duo), and find the models perform with vast variability. Under the single-task evaluation, o1 models achieve new SOTA and unlock challenges never solved by non-reasoning frontier models. But under the duo-task evaluation, o1 models perform worse than Claude 3.5, and consistently fail under specific test format.

We attempt to gain insights into o1 behaviors by deep diving into a few problems they succeed or fail at. We find the reasoning steps play critical role in both success and failure. Since reasoning tokens are invisible in OpenAI API, we share reasoning steps obtained from ChatGPT reeactment, i.e. feeding the identical prompt to ChatGPT. To minimize benchmark contamination, we only share test cases details, but do not reveal verbatim answers, only illustrate them in broad strokes.

The artifacts are on GitHub and Huggingface: single-task benchmark\citep{webapp1k-dataset}, dual-task benchmark\citep{webapp1kduo-dataset}, and the leaderboard\citep{webapp1k-leaderboard}. 

The rest of this report is organized as follows. Sec.~\ref{sec:webapp1k} presents results of single-task benchmark and how o1 models solve two hard problems. Sec.~\ref{sec:webapp1kduo} presents results of duo-task benchmark and how o1 models suffer in two testing scenarios. Sec.~\ref{sec:related} discusses related works. Sec.~\ref{sec:conclude} concludes and shares departing thoughts.
\section{Single-Task Benchmark}\label{sec:webapp1k}
We start with model performances on the WebApp1K benchmark. As illustrated in Tab.~\ref{tab:webapp1k_tests}, each challenge of the benchmark focuses on a single task described by two test cases, one success and one failure. The task is about completing an atomic action (e.g. submitting a form, retrieving all posts), involving user interactions and access to a mocked API. More details of the benchmark can be found at \citep{webapp1k-paper}.
\begin{table}[h!]
    \centering
    \begin{minipage}{0.48\textwidth}
        \centering
        \begin{tabular}{|l|}
            \hline
            \begin{minipage}{\dimexpr\textwidth-2\fboxsep-2\fboxrule}
                \vspace{2mm}
                \begingroup
                \renewcommand{\ttdefault}{pcr}
                \scriptsize
                \begin{verbatim}
...
import TaskA from './TaskA';

test("Success at task A", async () => {
  ...
  render(
    <MemoryRouter><TaskA /></MemoryRouter>
  );
  ...
}, 10000);
                \end{verbatim}
                \endgroup
            \end{minipage} \\
            \hline
        \end{tabular}
        \subcaption{Success Case for Task A}
    \end{minipage}
    \hspace{0.02\textwidth}
    \begin{minipage}{0.48\textwidth}
        \centering
        \begin{tabular}{|l|}
            \hline
            \begin{minipage}{\dimexpr\textwidth-2\fboxsep-2\fboxrule}
                \vspace{2mm}
                \begingroup
                \renewcommand{\ttdefault}{pcr}
                \scriptsize
                \begin{verbatim}
...
import TaskA from './TaskA';

test("Failure at task A", async () => {
  ...
  render(
    <MemoryRouter><TaskA /></MemoryRouter>
  );
  ...
}, 10000);
                \end{verbatim}
                \endgroup
            \end{minipage} \\
            \hline
        \end{tabular}
        \subcaption{Failure Case for Task A}
    \end{minipage}
    \caption{Illustration of WebApp1K Test Cases}
    \label{tab:webapp1k_tests}
\end{table}

The prompt is straightforward: we feed test files to the model, expecting it to generate code passing these tests.
\begin{align}
	&\text{Generate TaskA.js to pass the tests below: } \label{eq:prompt} &\\
	&\{Tab.~\ref{tab:webapp1k_tests} (a)\}\{Tab.~\ref{tab:webapp1k_tests} (b)\}.\text{ RETURN CODE ONLY.} \nonumber &
\end{align}

The resulting lines of code is typically between 40 and 50.
\subsection{Results}
Due to budget constraints, we only obtained $pass@1$ results for the o1 models. Nevertheless, as shown in Tab.~\ref{tab:webapp1k}, they demonstrate impressive performance, lifting SOTA by 7\%.
\begin{table}[h!]
\centering
\begin{tabular}{|l|c|}
\hline
\textbf{Model} & \textbf{pass@1} \\ \hline
o1-preview & 0.952 \\ \hline
o1-mini & 0.939 \\ \hline
gpt-4o-2024-08-06 & 0.885 \\ \hline
claude-3.5-sonnet & 0.881 \\ \hline
deepseek-v2.5 & 0.834 \\ \hline
mistral-large-2 & 0.780 \\ \hline
\end{tabular}
\caption{WebApp1K: pass@1 Results for Selected Models}
\label{tab:webapp1k}
\end{table}

As part of this achievement, the two o1 models unlock a total of 16 challenges never solved by previous non-reasoning models. Next, we pick two examples to illustrate how reasoning models solve them.
\subsection{Example One: Placeholder Text}
The first example is the \textit{postEditing} problem under the \textit{Social Media} category. In Tab.~\ref{tab:socialmedia}, we list the key steps to build up expectations of this problem. In particular, we highlight the step non-reasoning models overlooked.
\begin{table}[h!]
	\centering
	\begin{tabular}{|l|}
		\hline
		\begin{minipage}{\dimexpr\textwidth-2\fboxsep-2\fboxrule}
			\vspace{2mm}
			\begingroup
			\renewcommand{\ttdefault}{pcr}
			\scriptsize
			\begin{alltt}
				test('Test updating an existing post.', async () => {
				  fetchMock.post("/api/posts/1", 200);
				  ...
				  fireEvent.change(\textbf{screen.getByText('Edit')}, { target: { value: 'New content' } });
				  ...
				  fireEvent.click(screen.getByText('Save'));
				  ...
				  expect(fetchMock.calls("/api/comments").length).toBe(1);
				  expect(screen.getByText(/Comment added successfully/i)).toBeInTheDocument();
				}, 10000);
			\end{alltt}
			\endgroup
		\end{minipage} \\
		\hline
	\end{tabular}
	\caption{postEditing Problem}
	\label{tab:socialmedia}
\end{table}

First, the $fetchMock$ statement sets up a mocked API. Then, $fireEvent$ statements simulate user actions in two events: state change (value insertion) to an UI element carrying an \textbf{Edit} string, followed by a click event to an UI element carrying a Save string. Finally, $expect$ statements outline the expectations that the mocked API must be accessed exactly once, and the success response from the API must be present in the webpage.

For this problem, most non-reasoning models capture the semantics and deliver functioning code. Specifically, to support user actions, they implement a form element for user input, and a save button for the click event. 

However, they forget to explicitly attach the \textbf{Edit} string to the form element, without which $fireEvent$ cannot locate the correct element in the test webpage. There are two possible causes for the failure. First, the \textbf{Edit} token is synonymous with the purpose of the form element, which is also to edit. Second, the popular in-place editing implementation (prevelant in pretraining dataset) does not require an \textbf{Edit} string to state the purpose of the form element, which is overkill.

On the other hand, the o1 models stick to the requirement by attaching \textbf{Edit} to the form element as a placeholder text, via a $textarea$ attribute ($ref$ or $value$). Below is the ChatGPT reasoning chain, in which steps specifically reasoning \textbf{Edit} is blackened.
\begin{flushleft}
\hspace{1.5cm}Refining test details $\longrightarrow$ \textbf{Investigating the scripts} $\longrightarrow$ \\
\hspace{1.5cm}\textbf{Considering functionality} $\longrightarrow$ \textbf{Designing the component} $\longrightarrow$ \\
\hspace{1.5cm}Editing content $\longrightarrow$ \textbf{Refining selector logic} $\longrightarrow$ \\
\hspace{1.5cm}\textbf{Constructing a solution} $\longrightarrow$ \textbf{Setting up the interface} $\longrightarrow$ \\
\hspace{1.5cm}\textbf{Mapping out the test} $\longrightarrow$ \textbf{Trying another way} $\longrightarrow$ \\
\hspace{1.5cm}\textbf{Rendering editable text} $\longrightarrow$ Implementing the functionality $\longrightarrow$ \\
\hspace{1.5cm}\textbf{Mapping out test solutions} $\longrightarrow$ Revisiting test strategies $\longrightarrow$ \\
\hspace{1.5cm}\textbf{Weighing options} $\longrightarrow$ \textbf{Evaluating event handling} $\longrightarrow$ \\
\hspace{1.5cm}\textbf{Mulling over implementation} $\longrightarrow$ \textbf{Mapping the component} $\longrightarrow$ \\
\hspace{1.5cm}Testing with different methods $\longrightarrow$ \textbf{Formulating a solution} $\longrightarrow$ \\
\hspace{1.5cm}Managing content updates $\longrightarrow$ \textbf{Weighing options} $\longrightarrow$ \\
\hspace{1.5cm}Creating the component
\end{flushleft}
\subsection{Example Two: Frontend Validation vs Backend Validation}
The second example is the \textit{ticketSubmission} problem under the \textit{Customer Support} category. Tab.~\ref{tab:customersupport}, lists the key steps of the test setup and expectations. We blacken the step which trapped non-reasoning models.
\begin{table}[h!]
	\centering
	\begin{tabular}{|l|}
		\hline
		\begin{minipage}{\dimexpr\textwidth-2\fboxsep-2\fboxrule}
			\vspace{2mm}
			\begingroup
			\renewcommand{\ttdefault}{pcr}
			\scriptsize
			\begin{alltt}
				test('shows error when submitting a ticket with missing fields', async () => {
				  fetchMock.post('/api/tickets', { status: 400 });
				  ...
				  fireEvent.click(screen.getByText('Submit'));
				  ...
				  expect(fetchMock.calls('/api/tickets').length).toBe(1);
				  expect(\textbf{screen.getByText('Title is required')}).toBeInTheDocument();
				}, 10000);
			\end{alltt}
			\endgroup
		\end{minipage} \\
		\hline
	\end{tabular}
	\caption{ticketSubmission Problem}
	\label{tab:customersupport}
\end{table}

Similar to the same sequence in Tab.~\ref{tab:socialmedia}, the mocked API is first setup, followed by simulated user action, then expectations on API access and error message.

Again, non-reasoning models understand the semantics, write functioning code, but fail expectations. The root cause here is the string \textit{Title is required}, which is akin to a technique not requiring API access, aka frontend validation.  As a best practice (hence prevelance in pretraining dataset), frontend valiation is lightweight and fast, therefore preferred over backend validation. As such, all non-reasoning models are misled to implement frontend validation instead of expected behaviors which is backend validation.

On the other hand, o1 models discover the unpopular yet correct implementation: unconditionally visit the API, and output the \textit{Title is required} error message upon a 400 response. Below is the ChatGPT reasoning chain, in which steps reasoning the 400 response is blackened.
\begin{flushleft}
\hspace{1.5cm}Mapping out the component $\longrightarrow$ Setting up event handlers $\longrightarrow$ \\
\hspace{1.5cm}Setting up the form $\longrightarrow$ Writing test cases $\longrightarrow$ \\
\hspace{1.5cm}\textbf{Refining the approach} $\longrightarrow$ \textbf{Refining error handling} $\longrightarrow$ \\
\hspace{1.5cm}\textbf{Adjusting error handling} $\longrightarrow$ Adjusting code logic $\longrightarrow$ \\
\hspace{1.5cm}Updating JavaScript code
\end{flushleft}

The most crucial step here is \textbf{Refining the approach}. Below is its detailed wording.
\begin{flushleft} 
\hspace{1.5cm}I’m updating the code to ensure a fetch request is \textbf{always} sent, even without a title.\\
\hspace{1.5cm}The server will respond with a 400 status if the title is absent.
\end{flushleft}

Evidently, the step before it (Writing test cases) conducted certain verification, which leads the model to pivot to the right path.
\subsubsection{Counter Example}\label{sec:counterexample}
Unfortunatelly the reasoning models can also fall for the same trap. Below is a ChatGPT reasoning chain leading o1-preview to the faulty implementation like previous models. 
\begin{flushleft}
\hspace{1.5cm}Mapping out test strategy $\longrightarrow$ Setting up the test $\longrightarrow$ \\
\hspace{1.5cm}\textbf{Customer service improvement} $\longrightarrow$ Setting up for data $\longrightarrow$ \\
\hspace{1.5cm}Setting up the form $\longrightarrow$ \textbf{Verifying form submission} $\longrightarrow$ \\
\hspace{1.5cm}SHOWING ERRORS $\longrightarrow$ Refining the form handling 
\end{flushleft}

On a closer look, step \textbf{Customer service improvement} derails the model from backend validation to frontend validation.
\begin{flushleft} 
\hspace{1.5cm}I’m thinking about creating a TicketSubmission component with\\
\hspace{1.5cm}a 'Title' input and 'Submit' button. Submitting the form will trigger\\
\hspace{1.5cm}a POST request to '/api/tickets', validating the 'Title' field \textbf{before} submission.
\end{flushleft}

More interestingly, the step \textbf{Verifying form submission} does not correct the wrong direction, but solidify it.
\begin{flushleft} 
\hspace{1.5cm}I’m thinking about how the form ensures 'Title' must be filled.\\
\hspace{1.5cm}It sends a POST request \textbf{if} 'Title' is entered, showing success\\
\hspace{1.5cm}or 'Title is required' based on the response status.
\end{flushleft}

With these superficial clues, we speculate that the derailing is due to preemption of original expectations by model's inherent knowledge. The subsequent verification step is derived from neighboring steps already derailed, instead of orginal expectations only accessible from the input tokens. 
\section{Duo-Task Benchmark}\label{sec:webapp1kduo}
In light of o1 models' superb performance to saturate the single-task benchmark, we propose WebApp1K-Duo\citep{webapp1kduo-dataset}, a more difficult benchmark. Under each category of WebApp1K, we randomly pair up two atomic tasks into a duo task. The benchmark still consists of 1000 tasks, with 50 for each category. Models are challenged on both longer input, i.e. twice as many test cases, and longer output, i.e. more implementation in one module to meet all expectations.
\begin{table}[h!]
    \centering
    \begin{minipage}{0.48\textwidth}
        \centering
        \begin{tabular}{|l|}
            \hline
            \begin{minipage}{\dimexpr\textwidth-2\fboxsep-2\fboxrule}
                \vspace{2mm}
                \begingroup
                \renewcommand{\ttdefault}{pcr}
                \scriptsize
                \begin{alltt}
...
import \textbf{TaskA} from './TaskA_B';
import \textbf{TaskB} from './TaskA_B';

test("Success at task A", async () => {
  ...
  render(
    <MemoryRouter><\textbf{TaskA} /></MemoryRouter>
  );
  ...
}, 10000);

test("Failure at task A", async () => {
  ...
  render(
    <MemoryRouter><\textbf{TaskA} /></MemoryRouter>
  );
  ...
}, 10000);

test("Success at task B", async () => {
  ...
  render(
    <MemoryRouter><\textbf{TaskB} /></MemoryRouter>
  );
  ...
}, 10000);

test("Failure at task B", async () => {
  ...
  render(
    <MemoryRouter><\textbf{TaskB} /></MemoryRouter>
  );
  ...
}, 10000);
                \end{alltt}
                \endgroup
            \end{minipage} \\
            \hline
        \end{tabular}
        \subcaption{Raw Format}
    \end{minipage}
    \hspace{0.02\textwidth}
    \begin{minipage}{0.48\textwidth}
        \centering
        \begin{tabular}{|l|}
            \hline
            \begin{minipage}{\dimexpr\textwidth-2\fboxsep-2\fboxrule}
                \vspace{2mm}
                \begingroup
                \renewcommand{\ttdefault}{pcr}
                \scriptsize
                \begin{alltt}
...
...
import \textbf{App} from './TaskA_B';

test("Success at task A", async () => {
  ...
  render(
    <MemoryRouter><\textbf{App} /></MemoryRouter>
  );
  ...
}, 10000);

test("Failure at task A", async () => {
  ...
  render(
    <MemoryRouter><\textbf{App} /></MemoryRouter>
  );
  ...
}, 10000);

test("Success at task B", async () => {
  ...
  render(
    <MemoryRouter><\textbf{App} /></MemoryRouter>
  );
  ...
}, 10000);

test("Failure at task B", async () => {
  ...
  render(
    <MemoryRouter><\textbf{App} /></MemoryRouter>
  );
  ...
}, 10000);
                \end{alltt}
                \endgroup
            \end{minipage} \\
            \hline
        \end{tabular}
        \subcaption{Normalized Format}
    \end{minipage}
    \caption{Illustration of WebApp1K-Duo Test Cases}
    \label{tab:webapp1kduo}
\end{table}

WebApp1K-Duo is composed in two ways. The first way is shown in Tab.~\ref{tab:webapp1kduo} (a), in which the original export name of WebApp1K is preserved as is. The second way is shown in Tab.~\ref{tab:webapp1kduo} (b), where the export names are normalized to a unified name \textbf{App}.
\subsection{Results}
We collect $pass@1$ results under both raw and normalized formats. Unfortunately, o1 models' performances on the new benchmark are not impressive, falling behind other frontier models, especially Claude 3.5. 

As shown in Tab.~\ref{tab:webapp1kduo_raw}, all models struggle with the raw format (Tab.~\ref{tab:webapp1kduo} (a)). Most strikingly, o1 models fail all problems. We will try to find the root cause in Sec.~\ref{sec:duo_export}.
\begin{table}[h!]
\centering
\begin{tabular}{|l|c|}
\hline
\textbf{Model} & \textbf{pass@1} \\ \hline
claude-3-5-sonnet & 0.32 \\ \hline
chatgpt-4o-latest & 0.026 \\ \hline
deepseek-v2.5 & 0.02 \\ \hline
mistral-large-2 & 0.02 \\ \hline
o1-mini & 0 \\ \hline
o1-preview & 0 \\ \hline
\end{tabular}
\caption{WebApp1K-Duo Raw Format: pass@1 Results for Selected Models}
\label{tab:webapp1kduo_raw}
\end{table}

In Tab.~\ref{tab:webapp1kduo_normalized}, performance of all models are greatly improved under the intuitive normalized format (Tab.~\ref{tab:webapp1kduo} (a)). The SOTA is owned by Claude 3.5.
\begin{table}[h!]
\centering
\begin{tabular}{|l|c|}
\hline
\textbf{Model} & \textbf{pass@1} \\ \hline
claude-3-5-sonnet & 0.679 \\ \hline
o1-mini & 0.667 \\ \hline
o1-preview & 0.652 \\ \hline
chatgpt-4o-latest & 0.531 \\ \hline
deepseek-v2.5 & 0.49 \\ \hline
mistral-large-2 & 0.449 \\ \hline
\end{tabular}
\caption{WebApp1K-Duo Normalized Format: pass@1 Results for Selected Models}
\label{tab:webapp1kduo_normalized}
\end{table}
\subsection{Example One: Default Export vs Named Export}\label{sec:duo_export}
In the raw format illustrated in Tab.~\ref{tab:webapp1kduo} (a), there are two imports of different names, i.e. \textbf{TaskA} and \textbf{TaskB}. But they are actually default imports (without curly braces) which are name-agnostic. Also since only one default export is allowed per module, this format is in fact semantically equivalent to the normalized format in Tab.~\ref{tab:webapp1kduo} (b). Both formats demand the models to build a single module implementing all expectations, with a single default export. To help readers understand related concepts, we explain JavaScript export rules in Tab.~\ref{tab:exports}.
\begin{table}[h!]
\centering
\begin{tabular}{|l|l|l|}
\hline
                            & \textbf{Named Exports}                          & \textbf{Default Export}                       \\ \hline
\textbf{Purpose}            & Export multiple items from a module            & Export a single item from a module       \\ \hline
\textbf{Syntax}             & \texttt{export const x = ...;}                 & \texttt{export default ...;}                  \\ 
                            & \texttt{export function y() \{...\}}           &                                               \\ \hline
\textbf{Import Syntax}      & \texttt{import \{ x, y \} from}    & \texttt{import anyName from}      \\
                            &  \texttt{'./module';}                &   \texttt{'./module';}                                             \\ \hline
\textbf{Curly Braces}       & Required during import                         & Not required during import                    \\ \hline
\textbf{Import Naming}      & Must use the exact exported names              & Can be imported with any name                 \\ 
                            & (can use \texttt{as} to rename)                &                                               \\ \hline
\textbf{Multiplicity}  & Multiple named exports per module              & Only one default export per module            \\ \hline
\textbf{Use Case}    & Utility functions, constants, classes          & Main functionality of a module                \\ \hline
\textbf{Export Location}    & Anywhere in the module           & Bottom or after the main logic \\ \hline
\end{tabular}
\caption{Illustration of JavaScript Default Export in Comparison to Named Imports}
\label{tab:exports}
\end{table}

Tab.~\ref{tab:solutions_raw} collects different ways models cope with this challenge. Tab.~\ref{tab:solutions_raw} (d) is the only right answer, but also the least straightforward, challenging the intuition trap that two exports from two separate modules are needed. Both non-reasoning and reasoning models fall for the trap and attempt to split the implementation into two modules, (Tab.~\ref{tab:solutions_raw} (a), (b), (c)), resulting in very high failure rates.
\begin{table}[h!]
    \centering
    \begin{minipage}{0.48\textwidth}
        \centering
        \begin{tabular}{|l|}
            \hline
            \begin{minipage}{\dimexpr\textwidth-2\fboxsep-2\fboxrule}
                \vspace{2mm}
                \begingroup
                \renewcommand{\ttdefault}{pcr}
                \scriptsize
                \begin{verbatim}
function TaskA() {
  // Implementation of TaskA
}

function TaskB() {
  // Implementation of TaskB
}
export default TaskA;
export { TaskB };
                \end{verbatim}
                \endgroup
            \end{minipage} \\
            \hline
        \end{tabular}
        \subcaption{One Default Export and One Named Export}
    \end{minipage}
    \hspace{0.02\textwidth}
    \begin{minipage}{0.48\textwidth}
        \centering
        \begin{tabular}{|l|}
            \hline
            \begin{minipage}{\dimexpr\textwidth-2\fboxsep-2\fboxrule}
                \vspace{2mm}
                \begingroup
                \renewcommand{\ttdefault}{pcr}
                \scriptsize
                \begin{verbatim}
function TaskA() {
  // Implementation of TaskA
}

function TaskB() {
  // Implementation of TaskB
}

export { TaskA, TaskB };
                \end{verbatim}
                \endgroup
            \end{minipage} \\
            \hline
        \end{tabular}
        \subcaption{Two Named Exports}
    \end{minipage}

    \vspace{0.4cm}

    \begin{minipage}{0.48\textwidth}
        \centering
        \begin{tabular}{|l|}
            \hline
            \begin{minipage}{\dimexpr\textwidth-2\fboxsep-2\fboxrule}
                \vspace{2mm}
                \begingroup
                \renewcommand{\ttdefault}{pcr}
                \scriptsize
                \begin{verbatim}
function TaskA_or_B() {
  // Implementation of TaskA or TaskB
}

export default TaskA_or_B;
                \end{verbatim}
                \endgroup
            \end{minipage} \\
            \hline
        \end{tabular}
        \subcaption{Only One Task is Implemented and Exported}
    \end{minipage}
    \hspace{0.02\textwidth}
    \begin{minipage}{0.48\textwidth}
        \centering
        \begin{tabular}{|l|}
            \hline
            \begin{minipage}{\dimexpr\textwidth-2\fboxsep-2\fboxrule}
                \vspace{2mm}
                \begingroup
                \renewcommand{\ttdefault}{pcr}
                \scriptsize
                \begin{verbatim}
function TaskA_or_B() {
  // Implementation of both TaskA and TaskB
}

export default TaskA_or_B;
                \end{verbatim}
                \endgroup
            \end{minipage} \\
            \hline
        \end{tabular}
        \subcaption{Two Tasks Jointly Implemented and Exported}
    \end{minipage}
    \caption{Patterns to Address the WebApp1K-Duo Raw Format (Tab.~\ref{tab:webapp1kduo} (a))}
    \label{tab:solutions_raw}
\end{table}

Next, we try to understand why non-reasoning models occasionally succeed by following the pattern of Tab.~\ref{tab:solutions_raw} (d), but non-reasoning models never do so. We suspect that the normalized format (Tab.~\ref{tab:webapp1kduo} (b)) definitely dominates the pretraining/posttraining dataset, but does not exclude the raw format (Tab.~\ref{tab:webapp1kduo} (a)), as well as the matching solutions. This makes the success possible.

On the other hand, from the first reasoning step which often plays the role of planning, reasoning models commit to the wrong judgment, and do not get a chance to correct the course in subsequent steps. Below is the detailed wording of the first reasoning step from a ChatGPT reeactment.
\begin{flushleft} 
\hspace{1.5cm}To progress, the key task is creating components TaskA and TaskB in TaskA\_B.js \\
\hspace{1.5cm}to ensure all tests are successfully passed.
\end{flushleft}

Comparing to the mistakes made in Sec.~\ref{sec:counterexample}, the mistake in the above step covers a larger scope. It is reasonable to argue that mistakes made in large-scoped steps are more fatal and harder to correct.
\subsection{Example Two: Ignored Expectation}
We now try to study why o1 models perform worse than Claude 3.5 under the normazlied format. Tab.~\ref{tab:blogging} shows a problem solved by Claude 3.5, but failed by o1-preview.
\begin{table}[h!]
    \centering
    \begin{tabular}{|l|}
        \hline
        \begin{minipage}{\dimexpr\linewidth-2\fboxsep-2\fboxrule\relax}
            \vspace{2mm}
            \begingroup
            \renewcommand{\ttdefault}{pcr}
            \scriptsize
            \begin{alltt}
import App from './addComment_retrieveAllBlogPosts';
...
test('successfully adds a comment to a post', async () => {
  fetchMock.post('/api/comments', 200);
  ...
  expect(fetchMock.calls('/api/comments').length).toBe(1);
  expect(screen.getByText(/Comment added successfully/i)).toBeInTheDocument();
}, 10000);

test('fails to add a comment to a post', async () => {
  fetchMock.post('/api/comments', 500);
  ...
  expect(fetchMock.calls('/api/comments').length).toBe(1);
  expect(screen.getByText(/Failed to add comment/i)).toBeInTheDocument();
}, 10000);

test('Success: retrieve a list of all blog posts', async () => {
  fetchMock.get('/api/posts', { status: 200, body: [{ id: 1, title: 'First Post' }, 
                                                    { id: 2, title: 'Second Post' }] });
  ...
  expect(fetchMock.calls()).toHaveLength(1);
  expect(screen.getByText('First Post')).toBeInTheDocument();
  expect(screen.getByText('Second Post')).toBeInTheDocument();
}, 10000);

test('Failure: retrieve a list of blog posts with server error', async () => {
  fetchMock.get('/api/posts', { status: 500, body: { error: 'Internal Server Error' } });
  ...
  expect(fetchMock.calls()).toHaveLength(1);
  expect(\textbf{screen.getByText('Internal Server Error')}).toBeInTheDocument();
}, 10000);
            \end{alltt}
            \endgroup
        \end{minipage} \\
        \hline
    \end{tabular}
    \caption{addComment\_retrieveAllBlogPosts Problem}
    \label{tab:blogging}
\end{table}

Here, o1-preview passes all tests but the last one. The output code neither attempt to catch the 500 error nor print out the \textbf{Internal Server Error} string. The reasoning chain is normal, and no step specifically mentions the need to catch internal server errors.
\begin{flushleft} 
\hspace{1.5cm}Crafting the component $\longrightarrow$ Laying out the requirements $\longrightarrow$ \\
\hspace{1.5cm}Importing dependencies $\longrightarrow$ Breaking down the code $\longrightarrow$ \\
\hspace{1.5cm}Setting up the app $\longrightarrow$ Testing a post functionality $\longrightarrow$ \\
\hspace{1.5cm}Testing API integration
\end{flushleft}

Also o1-preview's inherent coding ability is solid, because it solves the retrieveAllBlogPosts problem when evaluated under the single-task benchmark. To this end, we suspect the root cause to be failure to pick up the expectation from input tokens, possibly due to length constraint. This mistake should be considered a matter of instruction following, which is applicable to both non-reasoning and reasoning models.
\section{Related Works}\label{sec:related}
The impressive achievements of reasoning models bulit on advancements from machine learning, reinforcement learning, and cognitive science. On the learning side, self-play fine-tuning allows models to generate their own data and iteratively refine their reasoning capabilities\citep{spft}. By engaging in self-play, models learn from successes and failures to convert weak performance into strong, well-aligned behavior\citep{sp-survey}. Self-taught reasoning methods use the model's own outputs to enable a bootstrapping process to improve future performance\citep{star}. This is evident in the development of self-taught reasoners, where models analyze outcomes of their reasoning chains\citep{quiet-star}. Reinforcement learning further augments this self-improvement process by allowing models to optimize their decision-making strategies via interaction with the running environment\citep{alphazero}.

On the inference side, chain-of-thought reasoning trains models to generate intermediate steps that mirror human-like thought processes\citep{cot-reasoning,verifystepbystep}. Inductive reasoning and hypothesis search techniques enable models to explore a space of possible outcomes, making it excel at abstract reasoning tasks\citep{hypothesissearch}. Advanced sampling methods, like repeated sampling and tree search, enhance the model's capacity to handle uncertainty\citep{fast-slow}. Together, these strategies provide a robust framework for models to perform nuanced and sophisticated reasoning in a wide variety of tasks\citep{mathword}.

On the evaluation side, more benchmarks have been proposed to focus on problem-solving capabilities in near-real-world environments. SWE-bench\citep{swebench} provides a comprehensive suite targeting core software engineering activities such as code generation, completion, error detection, and debugging. BFCL\citep{bfcl} assesses models' ability to generate accurate function calls, including prompt interpretation and argument handling. BIRD\citep{bird} evaluates models' proficiency in translating natural language queries into SQL codes. The Aider Leaderboard\citep{aider} ranks models based on their performance in real-world programming tasks such as bug fixing, refactoring, and code completion.
\section{Conclusions}\label{sec:conclude}
This report studies the latest reasoning models by OpenAI in the context of writing code to specific test expectations. We see both exciting and discouraging results, and share our investigations to gain more insights, especially how reasoning influence the outcome. We further argue that OpenAI's top-notch base model and SFT are equally important to the success of reasoning models. We believe that further advancements in these existing directions will continue to enhance reasoning models' performance, both amplifying strengths and mitigating weaknesses.

Below are our thoughts on next steps.
\begin{itemize}
\item We think the current SOTA of the duo-task benchmark (Tab.~\ref{tab:webapp1kduo_raw}) is a good milestone for hill climbing. So we do not plan to add more test cases until the next significant leap.
\item We will look deeper into error logs. But it would be quite surprising if we discover new error patterns besides those already identified\citep{webapp1k-insights}.
\item We will incorporate more frameworks (e.g. Vue) and languages (e.g. Python) to increase the benchmark coverage.
\end{itemize}
\bibliographystyle{plainnat}
\bibliography{ref}
\end{document}